\documentclass[twocolumn,showpacs,preprintnumbers,amsmath,amssymb]{revtex4}
\usepackage{graphicx}% Include figure files
\usepackage{dcolumn}% Align table columns on decimal point
\usepackage{bm}% bold math
\usepackage{color}
\usepackage{float}

\def\ed{\end{document}}

\def\sk{\smallskip}

\def\bg{\bigskip}

\def\la{\left(}
\def\rf{\right)}
\def\|{\'\i}
\def\beq{\begin{eqnarray}}
\def\eq{\end{eqnarray}}
\def\beqn{\begin{eqnarray*}}
\def\eqn{\end{eqnarray*}}
\def\nl{\noindent}

\def\ed{\end{document}}
\begin{document}

\title{Vector-bilepton Contribution to Four Lepton Production at the LHC}

\author{E. Ramirez Barreto}
\email{elmer@if.ufrj.br}
\affiliation{Centro de Ci\^encias Naturais e Humanas, UFABC \\
Santo Andr\'e, SP, Brazil}
\author{Y. A. Coutinho\footnote{Now at CERN, Switzerland}} 
\email{yara.amaral.coutinho@cern.ch}
\affiliation{
Instituto de F\'isica - Universidade Federal do Rio de Janeiro\\
Av. Athos da Silveira Ramos, 149 \\
Rio de Janeiro - RJ, 21941-972, Brazil
}
\author{J. S\'a Borges}
\email{jsaborges@uerj.br}
\affiliation{Instituto de F\'isica -- Universidade do Estado do Rio de Janeiro\\ 
Rua S\~ao Francisco Xavier, 524  \\
Rio de Janeiro - RJ, 20550-013, Brazil}
\date{\today}

\begin{abstract}
Some extensions of the standard model predict the existence of particles having two units of 
leptonic charge, known as bileptons. One of such 
models is based on the $SU(3)_c\times SU(3)_L\times U(1)_X$ symmetry group (3-3-1 model, for 
short). Our search uses the minimal  version of the 3-3-1 model, 
  having exotic charges for the quarks and new gauge bosons. This model  predicts the existence 
of bileptons as vector particles having  one ($V^{\pm}$)
 and two ($Y^{\pm\pm}$) units 
of electric charge.

We study the signatures for the  production of four leptons by considering the contribution of 
a pair of bileptons  in $p p $ collisions for 
three energy and luminosity regimes
 at the Large Hadron 
Collider (LHC).  We  present invariant mass and transverse momentum
distributions, the total cross section and we determine the expected number of events for 
each bilepton type.
Finally we analyze the LHC potential for discovering singly and doubly charged vector 
 bileptons at 95$\%$ C.L..   We conclude that the 
LHC collider can show a clear 
signature for the existence of bileptons as a signal of new physics. 

\end{abstract}

\maketitle

\section{Introduction}
A peculiar feature of the standard model (SM) is that none of the gauge bosons carry baryon 
or lepton number.
Many extensions of the SM predicted the existence of exotic particles that carry global
 quantum numbers, among
then leptoquarks and  bileptons. The bileptons are defined as bosons carrying  two units 
of lepton number and are
presents in $SU(15)$ grand unification theory and in the 3-3-1 model, for example.

In particular, the 3-3-1 model in its minimal version \cite{PIV, FRA} includes 
 singly ($V^\pm$) and doubly charged ($Y^{\pm\pm}$) bileptons  whereas   
  neutral and  single-charged bileptons  are present in
the model version with right-handed neutrinos \cite{RHN}. In addition to
these new gauge bosons, there are new quarks that carry two units
of lepton number, called  leptoquarks.

  It is interesting to study the prospects for the detection of
bileptons at linear and hadron colliders because its existence  leads to unique
experimental signature.  One expect that the LHC will spoil new physics and show up the 
existence of new particles. 
We guess that bilepton are among the
 new discoveries and we intend,
 in this work, to extend our analysis about the production of bileptons at the LHC.

It is known the effort of LHC experiments to search for new charged gauge boson. ATLAS and 
CMS  established model dependent bounds for $W^\prime$ mass, 
by analyzing lepton-antineutrino production from a sequential  $W^\prime$ particle,
 having the same coupling to fermions 
as the ordinary SM gauge boson \cite{WLI}. In the present  work we explore lepton-neutrino production from 
a singly charged  bilepton with peculiar properties and couplings 
for which the experimental mass limits do not apply. We also include the production
of a pair of same-sign lepton related to doubly charged bilepton contribution. There are no  
background for these
 processes however we consider the production of leptons from SM contribution  
 that can be misidentified with the signal and we show that they can be  easily eliminated 
by convenient cuts.

There are some previous works about the production of charged bileptons in the literature. Some authors established  model independent bounds for mass and 
couplings from low energy 
data and linear collider experiments. Using two versions of the 3-3-1  model, H. N. Long {\it et al.} \cite{LONG} studied the bilepton production 
in $e^+ e^-$ collider.  The contribution of doubly charged bileptons to four lepton 
production at linear collider was considered in \cite{LCO}.
For a detailed review about bileptons from an independent model approach see \cite{CUY}.

 From the same model B. Dion {\it et al.} \cite {DION} obtained the total cross section for the production of a pair of bilepton in 
hadron colliders, whereas in Ref. \cite {DUT} the authors  studied the production of just one bilepton associated with an exotic quark.
More recently, we consider two versions of this model  to analyze  singly and doubly 
 charged bilepton  pair production for LHC energies \cite{ELM1,ELM2,ELM3}.
It is shown that  the total cross section for vector bilepton production is three orders of magnitude larger than for scalar pair production 
for  $\sqrt s= 7$ TeV and $14$ TeV. For that reason  we do not include the scalar bilepton contribution in the present paper. 
 
  A recent  result was obtained by one of us for the decay of a 3-3-1  Higgs candidate into two  photons  where the  vector bilepton
play an important role \cite{HIGGS, ELM4}. Their analysis show that a bilepton  doublet with mass $\simeq  213$ 
 GeV and an ${\cal O}(10 \%) $ branching
ratio of the Higgs boson into invisible states can reasonably fit the available data. It was also shown that bileptons are 
associated with leptoquark production \cite{ELM5}.

Here we  analyze the contribution of bileptons to lepton production. We
 consider an unexplored bilepton mass domain accessible by the LHC 
(from $200$ to $700$ GeV)  by respecting peculiar relations between the gauge boson masses. 
  
We  consider the elementary process $q +\bar q$ 
where, apart from the standard model particles, the exotic quarks and the extra neutral 
gauge boson $Z^\prime$ contributions are taken into account.  
 In the section II
we give a brief review of the 3-3-1 model, section III is devoted to our results for the total cross section and 
 final lepton  distributions, 
and  section IV is devoted to our  conclusion.

\section{The 3-3-1 model}

The electric charge operator is defined  as:  
\begin{equation}
Q = T_3 + \beta \ T_8 + X I
\label {beta} 
\end{equation}
\noindent where $T_3$ and $ T_8$ are two of the eight generators satisfying the $SU(3)$ algebra
 $I$ is the unit matrix and $X$ denotes the $U(1)$ charge. Besides the ordinary standard model 
gauge bosons, the model predicts the existence of a neutral  $Z^\prime$,  double 
charged $Y^{\pm \pm}$ and single charged  $V^\pm$ gauge bosons.  
The charge operator determines how the fields are arranged in each representation and depends 
on the $\beta$ parameter.  
Among the possible choices, $\beta = - \sqrt 3$  \cite{PIV, FRA} corresponds to the minimal 
version of the model, whereas  $\beta = 1/ \sqrt 3$ leads to a model with right-handed neutrinos 
and no exotic charged fields \cite{RHN}.

In the minimal version of the model the left- and right-handed  lepton components  of each 
generation belong to triplet representation of  $SU(3)$.  
The procedure to cancel model anomalies imposes that quark families be assigned to 
different $SU(3)$ representation 
\cite{CAR}.
 Here we elect the left component of the first quark family to be accommodated 
in $SU(3)$ triplet and the second and third 
families ($m =2, 3$) to belong to the  
anti-triplet representation  as follows:
\begin{eqnarray}
&& Q_{1 L} =  \left( u_1 \ d_1 \  J_1
\right)_{L}^T \ \sim \left( {\bf 3}, 2/3 \right) \nonumber \\
&& Q_{m L} =  \left( d_m \  u_m \ j_m
\right)_{L}^T \ \sim \left({\bf 3^*}, -1/3 \right),
\end{eqnarray}
\nl the corresponding right handed component are:
\begin{eqnarray}
&& u_{a R}\ \sim \left({\bf 1}, 2/3 \right), \  d_{a R} \ \sim \left({\bf 1}, -1/3 \right) \nonumber \\
 && J_{ 1 R}\ \sim \left( {\bf 1}, 5/3 \right), \  j_{m R} \ \sim \left({\bf 1}, -4/3 \right), \nonumber
\end{eqnarray} 
\nl where $a = 1, 2, 3$ and $J_1$, $j_2$ and $j_3$ are exotic quarks with respectively 
$5/3$, $-4/3$ and $-4/3$  
 units of the positron charge ($e$). The numbers inside the  parentheses are the $SU(3)$  representation dimension
and the $X$ charge of each quark.

The Higgs structure necessary for symmetry breaking and that gives to quarks acceptable masses includes three triplets ($\eta, \rho$ and $\chi$) and a scalar 
($\sigma_2$)
in the sextet representation 
 that generates the correct lepton mass spectrum \cite{FOO}.
The neutral field  of each scalar multiplet develops non-zero vacuum expectation value ($v_\chi$, $v_\rho$, $v_\eta$, and $v_{\sigma_{2}}$) and the consistency of the model with the SM phenomenology is imposed by fixing a large scale for  $v_\chi$, responsible to give mass to the exotic particles 
 ($v_\chi \gg v_\rho, v_\eta, v_{\sigma_{2}}$), with $v_\rho^2 + v_\eta^2 + v^2_{\sigma_{2}}= 
v_W^2= \left( 246 \right)^2$ GeV$^2$.  

We call attention to the relation between $Z^\prime$, $V^\pm$ and $Y^{\pm\pm}$
 masses \cite{DION, DNG}:
\begin{equation}
\frac{M_{V}}{M_{{Z^{\prime}}}} \, =\, \frac {M_Y}{ M_{Z^\prime}} \, =\,  \frac{\sqrt{3 - 12 \sin^2 \theta_{_W}}}{ 2 \, \cos\theta_{_W}}. \label{essa}
\end{equation}

This  constraint respects the experimental bounds and it is equivalent to the $W$ to $Z$ 
masses relation in the SM. 
This ratio is $\simeq 0.3 $ for $\sin^2 \theta_{_W} =0.23$ \cite{PDG}, and so,
 $Z^{\prime}$ can decay into a bilepton pair.
\sk

\begin{widetext}
The  charged current interaction of leptons ($\ell$) with vector-bilepton  are  given by: 
\beq
{\cal L}^{CC}=-\frac{g}{\sqrt2}\sum_\ell \left[ \bar \ell^c \ \gamma^\mu\gamma^5\, \ell
\ {Y^{++}_\mu}  + \bar \ell^c \ \gamma^\mu\la 1 -\gamma^5 \rf \, \nu_\ell  
\ {V^{+}_\mu}\right] + h.c..
\eq
In the neutral gauge  sector, the  interactions of fermions $\Psi_f$ and bosons are described by the Lagrangian:
\begin{eqnarray}
{{\cal L}_{NC}}= \sum_{f} e q_f  \bar
\Psi_f\, \gamma^\mu  \Psi_f A_\mu - \frac{g}{2\, \cos{\theta_W}} \left[ \bar
\Psi_f\, \gamma^\mu\ (g_{V{_f}} - g_{A_f}\gamma^5)\ \Psi_f\, Z_\mu 
%\right.\nonumber 
%+ \left.
+ \bar\Psi_f\, \gamma^\mu\ (g^\prime_{V_f} - g^\prime_{A_i}\gamma^5)\ \Psi_f\, { Z_\mu^\prime} \right]
\end{eqnarray}
\noindent where  $e\,  q_f$ is the fermion electric charge and 
$ g_{V_f}$, $g_{A_f}$, $g^\prime_{V_f}$ and $ g^\prime_{A_f}$ are the fermion vector and axial-vector couplings 
with $Z$ and $Z^{\prime}$ respectively, displayed in Table I. 
The trilinear couplings from the self-interactions of gauge fields are shown in  Table II.

Finally, the couplings of ordinary to exotic quarks are driven by  charged bilepton as follows:  
\begin{eqnarray}
{\cal L}^{CC}= - \frac{g}{2 \sqrt 2} \left[ \bar u \, \gamma^\mu \la 1 - \gamma^5 \rf \, \la { {\cal U}_{21} \ j_2 +
{\cal U}_{31} \ j_3  } \rf  + \bar J_1 \gamma^\mu \la 1 - \gamma^5 \rf \ {\cal V}_{11} \ d \right] Y^{++}_\mu \nonumber \\ +
\left[\bar d \gamma^\mu \la 1 - \gamma^5 \rf  \la {\cal V}_{21} \,\,j_2 + 
{\cal V}_{31} \,\,j_3 \rf + \bar J_1 \gamma^\mu \la 1 - \gamma^5 \rf \,\, {\cal U}_{11} \,\,u \right] V^+_\mu + h.c. 
\end{eqnarray}
\noindent where ${\cal V}_{ij}$, ${\cal U}_{ij}$  are mixing matrices elements.
\end{widetext}

From this expression, and considering the  leptonic number conservation, we conclude that  
the exotic quarks carries
 two units of leptonic quantum number and  so they are a class of leptoquarks.  
 
\section{Results}

Let us first consider the experimental bounds on bilepton masses. As it was explained in the 
introduction, 
the experimental limits for $W^\prime$ mass established by ATLAS and CMS collaborations do
 not apply to the bilepton $V^\pm$ mass \cite{ATLAS}, because this bilepton couples with 
exotic quarks 
in addition to the ordinary quarks.
On the other hand, a bound on the  doubly charged bilepton mass of $510$ 
GeV was obtained from LEP data when including exotic contributions to $\mu^+ \mu^- $ 
and $\tau^+ 
\tau^-$ production; this bound increases to $740$ GeV when lepton-flavour violating charged
lepton decay data are included \cite{TRU}. A model independent and more recent analysis shows 
that a bilepton mass values around $500$ GeV is at the limit of the exclusion region from 
LEP data for an specific range of double charged bilepton-lepton coupling \cite{GREG}. 
The relation between 
bileptons to  $Z^\prime$ masses 
gives  $Z^\prime$ mass values used in the present work ($\simeq 1.1, 1.8 $ and $2.6$ TeV) 
not excluded by the experimental bounds that are  model dependent  
\cite{CMS}.

On the theoretical point of view, a small bilepton mass (around $500$ GeV) is convenient for 
respecting the constraint on the three measurable quantities, called $S$,  $T$, and $U$, 
that parameterize potential new physics contributions to electroweak radiative corrections. 
Another important issue related to the gauge boson mass in the $3-3-1$ model is 
a Landau pole that appears when the ratio of $SU(3)_L$ to $U(1)_X$ coupling constants becomes 
infinite at a finite energy scale. 
To avoid this critical situation, the $Z^\prime$ must be kept below $4$ TeV \cite{ALEX}.
From these remarks, we adopt three values for the singly charged bilepton mass namely: 
 $300$, $500$ and $700$ GeV, corresponding M$_{Z^\prime}= 1.0, 1.7$ and $2.3$ TeV respectively.

Next we consider bilepton contribution for lepton-neutrino production. For the three values of 
bilepton mass, 
the corresponding widths are  $0.8 $, $ 1.4 $ and $ 2.0 $ GeV.
We do an important choice in order  to avoid $V^\pm$ decaying into exotic quarks by fixing, 
 in our calculation, the 
exotic quark mass equal to $1$ TeV. This way, $V^\pm$ decays only  into the three 
lepton flavors with  the same  branching ratios.

In our procedure we consider the production of two pairs of lepton-neutrino in proton-proton 
collision from the process below:
$$q + \bar q  \rightarrow \gamma,\  Z, \ Z^\prime \rightarrow  V^- + V^+ 
\rightarrow \{\ell + \nu_\ell\} + \{\bar \ell^\prime + \bar \nu_{\ell^\prime}\}, $$ 
\nl where $\ell$ and $\ell^\prime $  stands for electron or  muon and the braces indicate 
the particles corresponding to the decay of each charged gauge boson.

We calculate the total cross section and we generate the final state events by using  
the CompHep package 
\cite{HEP}. On the set of events previously generated,
we apply  
 convenient cuts for the detector 
acceptance, and kinematic cuts for final leptons: 
$$\vert \eta \vert \leq 3.0,\  p_{{T}} > 20 \ {\mbox{GeV\  and}} \ \not\!\!E_T > 20 \,  {\mbox {GeV}}.$$
From  the selected events and using    
MadGraph/MadAnalysis \cite{MAD} we obtain some distributions which allow us to compare  the signal from the SM process
$$q + \bar q  \rightarrow \gamma,\  Z  \rightarrow  W^- + W^+ \rightarrow \{\ell + \bar\nu_\ell\} + \{\bar \ell^\prime + \nu_{\ell^\prime}\},
 $$ 
and 
$$ q + \bar q \rightarrow  t  + \, \bar t 
\rightarrow W^+ + j +  W^- +  j  \rightarrow \{\ell  \bar +
\nu_{\ell} \} +  \{\bar \ell^\prime + \nu_{\ell^\prime} \}  + j + j, $$
\nl where $\ell$ and $\ell^\prime $ stands for electron or muon, the braces indicate 
the particles corresponding to the decay of each charged gauge boson and $j$ is an hadronic jet.

\nl In the Figure 1, we  presents our results for the lepton-neutrino  invariant mass distribution (up)  and lepton transverse momentum  distribution (down) at
 $\sqrt s = 7$ TeV. In the invariant mass distribution plot, we are including the signal  from the bilepton production plus the SM background. 
From this figure one observes that $W^+ \, W^-$ contribution 
is smaller than the three considered bilepton contributions 
for lepton invariant mass from $600$ GeV. On the other hand the $t$ $\bar t$ 
channel contributes less than bileptons for $200$ GeV dilepton invariant mass  and its 
contribution becomes even less important for larger dilepton  masses. 
One also observes that  when the final lepton comes from a bilepton,
its  transverse momentum distribution is smeared in a wide range of $p_T$ in contrast 
with the SM final lepton $p_T$ distribution.
These plots clearly show that it is possible to completely isolate the SM contributions 
by applying  a convenient cut on the lepton-antineutrino 
final state invariant mass and on the charged lepton transverse momentum.  

Finally, from the calculated cross section and considering $5$ fb$^{-1}$ 
integrated luminosity, one  expect 
 $3500$, $125$ and $8$  events {\it per} year   
 for $M_V = 300, 500$ and $700$ GeV respectively.

\begin{figure}[ht]
\begin{center}
\includegraphics[height=.25\textheight]{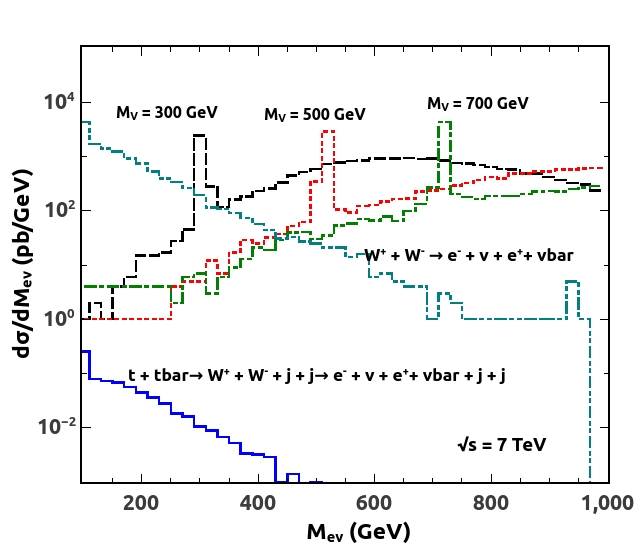}
\includegraphics[height=.25\textheight]{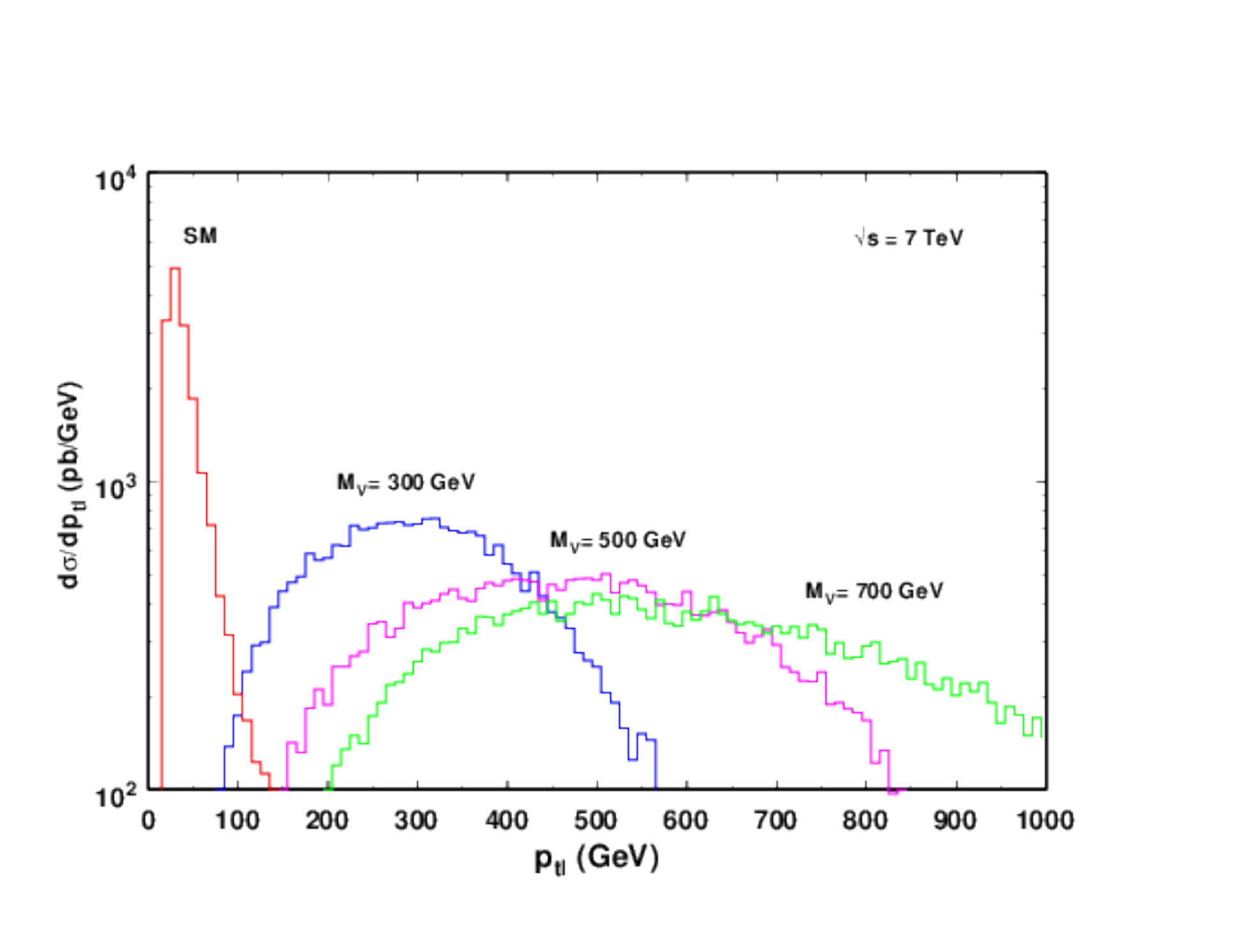}
\end{center} 
\caption{Lepton-neutrino and lepton-antineutrino (SM backgrounds) 
invariant mass distributions (up), and 3-3-1 and SM lepton transverse momentum distributions
(down), for three values of the  singly charged bilepton mass  at $\sqrt s= 7$ TeV.}  
\end{figure}

In the following we consider the contribution of the doubly charged bilepton for
 the production of two pairs of equal sign leptons:
$$q + \bar q  \longrightarrow \gamma, \  Z, \ Z^\prime \longrightarrow  Y^{--} + Y^{++} 
\longrightarrow \{\ell + \ell\} + 
\{\bar \ell^\prime + \bar \ell^\prime\}, $$ 
\nl where $\ell$ and $\ell^\prime$ stands for electron or muon and  the braces indicate the particles corresponding 
to the decay of each charged gauge boson.

This is an interesting process because it can encode  a non conservation of lepton number. 
This violation is much more clear in this case than for 
for singly charged induced process, where it can also occur.

In the present analysis we  respect the relation between gauge boson masses that characterizes 
the minimal version of the 3-3-1 model (Eq. \ref{essa}), 
 by selecting  three values for the bilepton mass 
($300, 500$ and $700$ GeV)
\, corresponding respectively to $M_{Z^\prime} \simeq 1.1, 1.8 $ and $2.6$ TeV. 

In our calculation, exotic quark t-channel exchange is taken into account to guarantee that
 elementary processes cross section 
does  respect unitarity. On the other hand,  we fix exotic quark masses equal to $1$ TeV to 
avoid a large jet 
production from bilepton decaying into exotic plus ordinary quark. 
This way we get equal branching fraction for bilepton ($M_Y = 300, 500$ and $700$ GeV) decaying
 into  same sign leptons  
with   widths  $\Gamma_Y = 2.5,\  4.2 $ and $ 5.9 $ GeV. 

We follow the same procedure as before  by adopting the kinematic cuts for final leptons:
$$ \vert \eta \vert \leq 3.0, \, p_T > 20 \, \mbox{GeV}, \, \, m_{\ell \ell} \, \, \, \mbox{and} \,\,  m_{\ell^\prime  \ell^\prime} > 50 \, \mbox{GeV}, $$
\nl to obtain $765,\ 41$ and $2$  events {\it per} year for the select masses at $\sqrt s = 7$ TeV and $5$ fb$^{-1}$ integrated luminosity. 

Figure 2 shows the results for equal-sign leptons invariant mass (up)  and 
lepton transverse momentum distributions (down). 
One observes the peak related to the resonance corresponding to three bilepton masses and their very narrow widths. Besides, 
lepton transverse momentum  distribution is similar to the case where it is produced from a single 
charged bilepton.

\begin{figure}[ht]
\begin{center}
\includegraphics[height=.3\textheight]{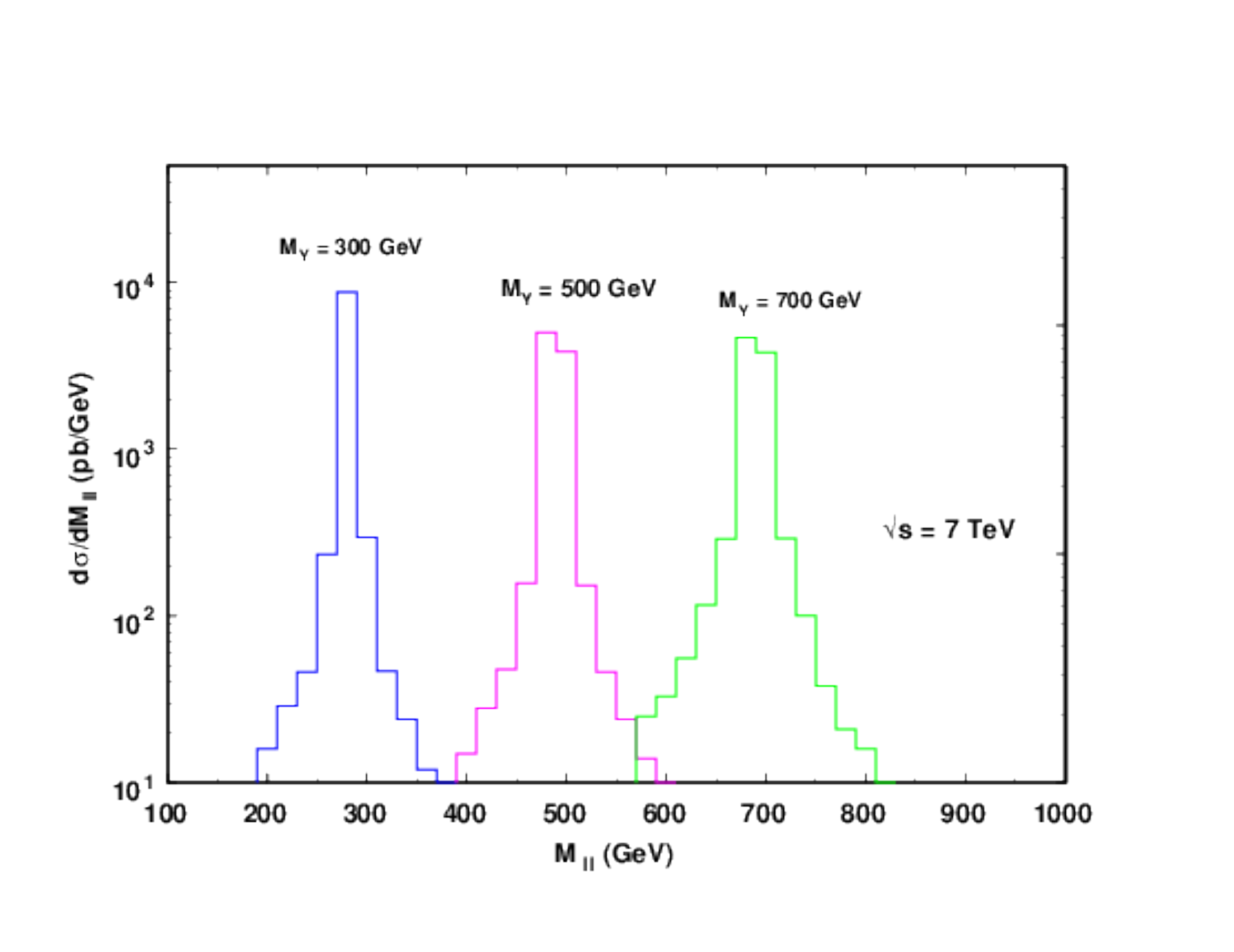}
\includegraphics[height=.3\textheight]{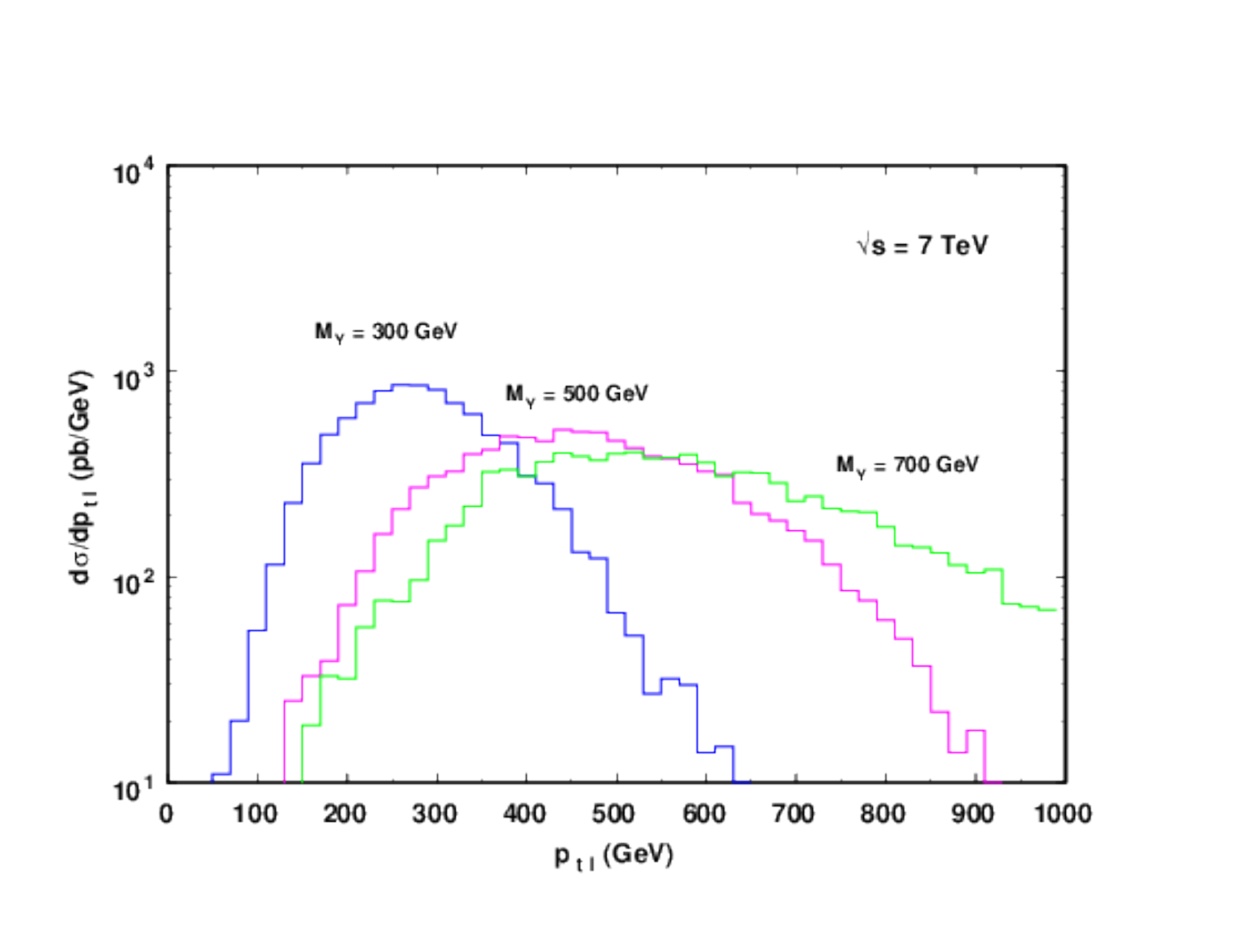}
\end{center}
\caption{Same sign lepton invariant mass  distribution (up)  and lepton  transverse momentum  
distribution (down)  for three values of the  doubly charged bilepton mass  at $\sqrt s= 7$ TeV.} 
\end{figure}

%\begin{figure}[ht]
%\begin{center}
%\includegraphics[height=.2\textheight]{Figure5}
%\end{center}
%\caption{Electron-neutrino invariant mass distribution from the SM background ($t \, \bar t$  and $W^+ \, W^-$) and from the 3-3-1 process ($V^+ \, V^-$ 
%considering 
%three bileptons masses, $M_V = 300, 500,$  and $700$ GeV) at $\sqrt s = 7$ TeV.} 
%\end{figure}

We have analyzed the LHC potential for discovering singly and doubly charged vector bileptons at
 $95\%$ C.L. and   $66\%$ for electron channel efficiency \, $\epsilon$, \  defined as
$\displaystyle{{\cal L}_{int} \, =\, 5 / \left[ \epsilon (m_V)\ \sigma_{tot} \, (m_V) \right]}$.

 Our results  in the Figure 4 show the calculated values of integrated luminosity as a 
function of bilepton mass. 
First it is shown in Figure 4 (up) that  ${\cal L} \simeq 5$ fb$^{-1}$  is sufficient
for discovering a  singly charged bilepton with $M_V$ up to $700$ GeV  at $\sqrt s = 7$ TeV.
For the same bilepton mass and higher energies ( $\sqrt s = 8, \ 14$ TeV) 
the integrated luminosities of $ 1$ and $ 0.1$ fb$^{-1}$ are required.

The LHC potential for discovering a doubly charged bilepton is represented 
in the Figure 4 (down) where one realize that these bileptons can be found for the same luminosity referred  before with $M_Y \simeq M_V - 100$ GeV.
We extended our analysis below $300$ GeV in order to consider the bilepton mass used in \cite{ELM4}.

The potential for discovering doubly charged vector bileptons
 at $\sqrt s = 7$ and  $ 14$ TeV was also obtained in \cite{AND}. In that paper,  the authors clearly violates  the constraint expressed in
 Eq. \ref{essa} by combining four bilepton mass values with a fixed   $Z^\prime$ mass equal to  $1$ TeV. Another 
issue that deserves a comment is the choice made by the authors exploring low exotic quark  masses ($< 1$ TeV), that gives an uncontrolled jet production rate.

\section{Conclusion}
The LHC at CERN opened the possibility to explore an energy regime where the standard model of electroweak interactions have not yet been tested. 
At these energies one expect that new resonances, associated with the existence of
extra gauge bosons like $W^\prime$ and/or $Z^\prime$, can be produced. These particles are predicted in some SM extensions or alternative models 
such as the 3-3-1 model studied in the present paper.  

The particle content of the minimal version of this  model includes scalars and gauge boson carrying two units of leptonic charge ($L$) , called bileptons, and 
new quarks (leptoquarks)  with exotic electric charges  ($4/3 e, \, 5/3 e$) leading to rich  phenomenological consequences.  In particular, this 
paper focus on lepton number violation in the production of one lepton pair ($L=+2$)  and a anti-lepton pair ($L=-2$) induced by bileptons.

An important characteristic of this model keeps some similarity with the SM one. We are referring to the relation between charged and neutral gauge boson 
 masses that in the SM is $M_W= \cos \theta_W \times M_Z$. Our calculation takes into account  the constraint expressed in the Eq. \ref {essa}  
 in contrast with \cite{AND}, where the authors  combined different charged bilepton mass with only one extra neutral gauge boson mass.

We perform the total cross section calculation using the 
CompHep and MadGraph packages to generate events that for ${\cal L} = 5$ fb$^{-1}$ annual integrated luminosity results in 
 a considerable number of leptons. 

The observation of a transverse lepton momentum distribution smeared  and locate at large $p_T$ values  is in contrast with the shape 
corresponding to the misidentified leptons from the SM resonance.  Moreover the invariant mass  distribution is shifted to large values in contrast with he SM 
distribution allowing to select the signal by a convenient cut. 

We analyzed the LHC discovery potential as a function of bilepton mass showing that  an integrated 
luminosity of order of $5$ fb$^{-1}$  is enough
for discovering a bilepton with $M_V \simeq 700$ GeV which implies 
that such signal can be observed for LHC at $7$ or $8$ TeV, whereas for the LHC running at $14$ TeV one needs lower luminosity. 

From our study in this specific model we conclude that the LHC is capable to show up signals for new particles existence related to new phenomena 
including lepton number violation.

\begin{figure}[ht]
\begin{center}
\includegraphics[height=.26\textheight]{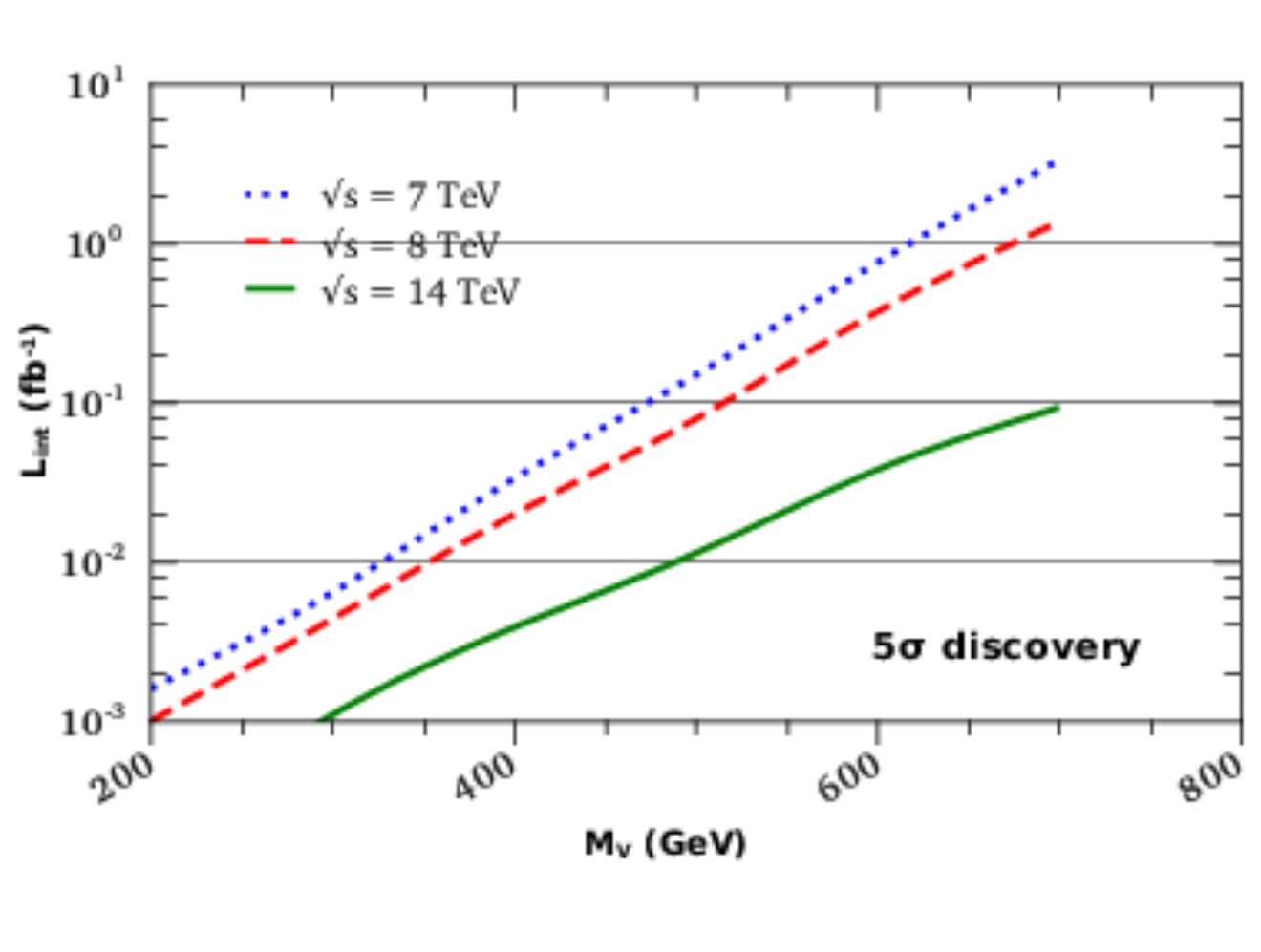}
\includegraphics[height=.27\textheight]{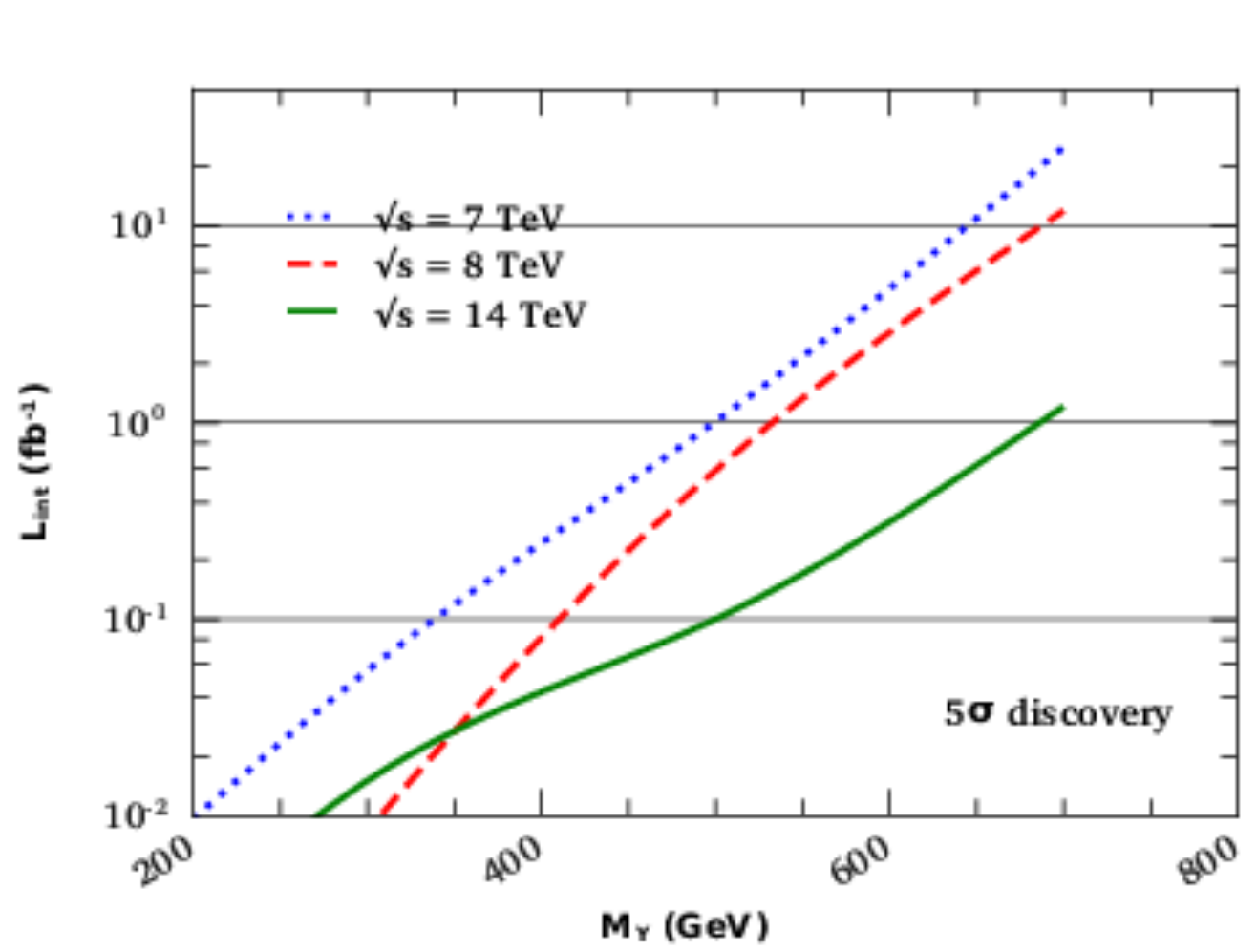}
\end{center}
\caption{Minimal integrated luminosity needed
for a $5 \sigma$ singly charged (up) and doubly charged (down) bilepton discovery at the LHC.} 
\end{figure}

\begin{table}[ht]\label{sagui}
\begin{footnotesize}
\begin{center}
\begin{tabular}{||c|c|c||}
     \hline
\hline
&    &          \\ 
&  Vector Couplings & Axial-Vector Couplings    \\
&    &          \\ \hline
\hline
&    &      \\
$Z \bar u_i u_i$ & $\displaystyle{\frac{1}{2}-\frac{4\, s^2_{_W}}{3}}$ &
$\displaystyle{\frac{1}{2}}$  \\
&      &   \\ \hline
\hline
&       &   \\
$Z \bar d_j d_j$  &   $\displaystyle{-\frac{1}{2}+\frac{2\, s^2_{_W}}{3}}$  &
$\displaystyle{-\frac{1}{2}}$  \\
&     &     \\  \hline
\hline
&     &     \\
$Z^{\prime} \bar u_i u_i$ & $\displaystyle{\frac{1-6\, s^2_{_W}-{\cal U}^*_{ii} {\cal U}_{ii} \, c^2_{_W}}{2\,\sqrt{3}\, r}}$
& $\displaystyle{\frac{1+2\,s^2_{_W}+{\cal U}^*_{ii} {\cal U}_{ii}\, c^2_{_W}}{2\,\sqrt{3}\, r}}$  \\
&    &    \\ \hline
\hline
&     &        \\
$Z^{\prime} \bar d_j d_j$ &  
$\displaystyle{\frac{1-{\cal V}^*_{jj} {\cal V}_{jj} \, c^2_{_W}}{2\,\sqrt{3}\, r}}$  &  $\displaystyle{\frac{  r^2\, +\, {\cal V}^*_{jj} {\cal V}_{jj} \,c^2_{_W}}{2\,\sqrt{3}\, r}}$  \\
&      &      \\ \hline
\hline
\end{tabular}
\end{center}
\end{footnotesize}
\caption{The $Z$ and $Z^{\prime}$ vector and axial-vector couplings to quarks ($u_1= u, u_2= c, u_3= t$, and $d_1= d, d_2= s, d_3= b$), 
${\cal U}_{ii}$ and ${\cal V}_{jj}$ are $\cal U$ and $\cal V$ diagonal mixing  matrix elements, with $s_{_W} = \sin \theta_W $, $c_{_W} = \cos \theta_W $  and  $r = \sqrt{1-4\,s^2_{_W}}$.}
\end{table}

\begin{table}[h]\label{chihuahua}
\begin{footnotesize}
\begin{center}
\begin{tabular}{||c|c|c|c||}
     \hline
\hline
    & & &          \\ 
  Vertex & $\gamma Y^{++} Y ^{--} $  & $Z Y^{++} Y ^{--} $ & $Z^{\prime} Y^{++} Y ^{--} $\\
    &  & &    \\ \hline
 &  & &  \\ 
Coupling  & $2 e$ &  
   $\displaystyle{\frac{r^2}{2 s_{_W}  c_{_W}}}$ 
&     $\displaystyle{\frac{\sqrt{3} }{2 s_{_W}  c_{_W}} \, \frac{r}{\sqrt{1-2 s^2_{_W}} }}$ \\
&     & &     \\ \hline
\hline
\end{tabular}
\end{center}
\end{footnotesize}
\caption{Couplings of neutral gauge bosons with  vector-bilepton $Y^{\pm \pm}$, with $s_{_W} = \sin \theta_W $, $c_{_W} = \cos \theta_W $ and  $r = \sqrt{1-4\,s^2_{_W}}$.}

\end{table}
\newpage

\nl {\Large \bf Acknowledgments}

E. Ramirez Barreto, Y. A. Coutinho and J. S\'a Borges thank Fapesp, Faperj and  CNPq for financial support respectively. 
\bg

\end{document}